\documentclass[preprint,pre,showpacs,groupedaddress]{revtex4}
\usepackage{amsmath}
\usepackage{graphicx}
\usepackage{amssymb}
\usepackage{bm}
\usepackage{epsfig}
\usepackage{amsopn}
\usepackage{natbib}
\usepackage{times}
\begin{document}
\title{Methods of exploring energy diffusion in lattices with finite temperature}
\author{Ping Hwang$^1$}
\author{Hong Zhao$^{1,2,}$}
\email{zhaoh@xmu.edu.cn}
\affiliation{$^1$Department of Physics, Xiamen University, Xiamen 361005,China\\
$^2$State Key Laboratory for Nonlinear Mechanics, Institute of Mechanics,
Chinese Academy of Sciences, Beijing, 100080, China}

\begin{abstract}
We discuss two methods for exploring energy diffusion in lattices
with finite temperature in this paper. The first one is the energy-kick
(EK) method. To apply this method, one adds an external energy kick
to a particle in the lattice, and tracks its evolution by evolving
the kicked system. The second one is the fluctuation-correlation (FC)
method. This method is presented recently by one of the present authors
{[}Zhao, Phys. Rev. Lett. \textbf{86}, 11003 (2006)]. In present paper,
the formula for calculating the probability density function (PDF)
using the canonical ensemble is slightly revised and extended to the
microcanonical ensemble. To apply the FC method, one tracks the motion
of the energy initially localized at a small region by a properly
constructed correlation function of energy fluctuations. Both methods
can obtain a PDF of energy diffusion. However, we show that the FC
method has advantages over the EK method theoretically and technically.
Theoretically, the PDF obtained by the FC method reveals the diffusion
processes of the inner energy while the PDF obtained by the EK method
represents that of the kick energy. The diffusion processes of the
inner energy and the external energy added to the system, i.e., the
kick energy, may be different quantitatively and even qualitatively
depending on models. To show these facts, we study not only the equilibrium
systems but also the stationary nonequilibrium systems. Examples showing
that the inner energy and the kick energy may have different diffusion
behavior are reported in both cases. Technically, since applying the
energy fluctuations of particles in the system, a set of independent
realizations to the ensemble average can be achieved by one round
of evolution of the system when applying the FC method. This advantage
enables us to study the long-time diffusion processes in large-scale
systems and thus avoids the finite-time effect.
\end{abstract}

\pacs{05.60.-k, 44.10.+i}

\maketitle

\section{introduction}

Diffusion is one of the most important types of motion in nature.
According to the time-dependent behavior, $\left\langle r^{2}(t)\right\rangle \sim t^{\alpha}$,
of the mean square displacement of a conserved quantity, it is classified
as subdiffusion ($\alpha<1$), normal diffusion ($\alpha=1$) and
superdiffusion ($\alpha>1$) \cite{metzler2000,zaslavsky2002}. Among
various quantities, the diffusion of particle and the diffusion of
energy have particular importance both for theoretical studies and
applications. The diffusion of particles represents the transport
behavior of mass, while the diffusion of energy determines the transport
of heat. In studying the particle diffusion, one can directly calculate
the PDF by tracking the motion of particles. This method has been
widely used \cite{perrin1909,marrero1972}. However, this method
is not available for energy diffusion, because a particle can be tagged
and tracked while energy may not be. Therefore, developing methods
to track energy diffusion is highly desirable. This is particularly
important for a lattice system since particles in the system oscillate
around their stationary positions and thus is not a diffusion process.

The key task of characterizing a diffusion process is to obtain the
probability density function (PDF) of the related quantity, by which
one can calculate the time-dependent behavior of the mean square displacement
and thereby obtain the diffusion exponent $\alpha$. This exponent
is not only served to classify diffusion types but also to check certain
theoretical predictions. In the past decade, the heat conduction problem
in low-dimensional systems has attracted intensive attentions \cite{livi2003,lepri2003}.
It is found that some one-dimensional lattices may show anomalous
heat conduction behavior \cite{rieder1967,lepri1997,hatano1999,lepri2000,dhar2001}.
The thermoconductivity $\kappa$ in such a system depends on the lattice
size $N$ as $\kappa\sim N^{\beta}$. While some other one-dimensional
lattices show normal heat conduction behavior \cite{casati1984,hu1998,hu2000,Giardina2000,Gendelman2000}.
In recent years, several groups investigated the relation between
the exponents $\alpha$ and $\beta$ \cite{libw2003,denisov2003,cipriani2005,libw2005,zhao2006,delfini2007},
and derived equations to describe their relationship. Establishing
deterministic relation between the two exponents has theoretical importance.
Actually, this attempt is stimulated by the celebrated equation obtained
by Einstein \cite{einstein1905} who derived the deterministic relation
between the diffusion coefficient and the coefficient of viscosity,
correlating the two different irreversible processes. However, up
to now, different relationship equations of $\alpha$ and $\beta$
coexist \cite{denisov2003,libw2003}. Which one describes the correct
relationship, and even whether the correct one has been found, are
still open problems. To answer these questions, the exponents $\alpha$
and $\beta$ should be calculated with sufficient precision after
getting rid of the size effect or finite-time effect. As shown in
Ref. \cite{mai2007,delfini199401comm,dhar2008Re}, the size effect
of $\beta$ disappears in the FPU-$\beta$ lattices until $N\sim10^{3}-10^{4}$.
This fact reminds one that the diffusion time should be long enough
to avoid the finite-time effect on the exponent $\alpha$.

Two different methods have been presented for calculating the PDF
of energy diffusion. The first one is a straightforward method \cite{zavt1993,arevalo2003,libw2005,cipriani2005,delfini2007,kopidakis2008}.
The idea is to add a high-energy kick to a particle at a fixed position
and then, after a period and at other positions, calculate the ensemble
average of the difference between the current energy density and the
average energy density before the kick. It is clear that the ensemble
average represents the amount of kick energy transported to these
positions. We call it the energy-kick (EK) method. The EK method indeed
manifests the diffusion of the kick energy. The second way is presented
by one of the present authors, H. Zhao, recently \cite{zhao2006}.
The basic idea is to track the motion of a part of energy by investigating
a properly constructed correlation function of energy fluctuations.
We call it the fluctuation-correlation (FC) method. Unlike the EK
method, it studies the energy fluctuations of particles in the system.
Some groups have realized that this method is more reasonable than
the EK method \cite{delfini2007,dhar2008}.

The purpose of this paper is, on one hand, to make a detailed comparison
between the EK method and the FC method. We argue that the two methods
indeed explore different diffusion processes. That is to say, the
EK method describes the relaxation process of the kick energy or the
diffusion behavior of the energy externally added to the system, while
the FC method represents the diffusion behavior of the inner energy
or the intrinsic diffusion behavior of the energy in the system. In
certain systems, the memory of the history will be lost quickly with
time evolution. In this case, the diffusion of the kick energy and
the inner energy has no qualitative difference, though the PDFs obtained
by the two method still have quantitative difference since the diffusion
of the kick energy depends on the amplitude of kicks. In other systems,
the memory of the history is not totally lost. In this case, the kick
energy and the inner energy may include different information. For
most lattices in equilibrium with uniform temperature, the kick energy
and the inner energy diffuse in a symmetric way, and the PDFs obtained
by both methods are qualitatively similar. These lattices usually
have symmetric interaction potentials. While for lattices with asymmetric
interaction potentials, the inner energy and the kick energy may diffuse
in different ways. The PDFs obtained by the FC method are symmetric
and those obtained by the EK method may be asymmetric. In nonequilibrium
lattices with stationary temperature gradients, even for those lattices
with symmetric interaction potentials, the PDFs obtained by the two
methods may be different qualitatively. Therefore, we consider not
only the energy diffusion in equilibrium systems but also in nonequilibrium
systems.

On the other hand, we further explain the rationale of the FC method
and describe the technical details for applying this method. We present
formulae for the canonical ensemble and the microcanonical ensemble
respectively to calculate PDFs of energy diffusion. A formula suitable
for the canonical ensemble has been presented in \cite{zhao2006}.
The one presented here is slightly different from the old one. We
explain that the new formula is more reasonable. Also we emphasize
that the FC method has great technique advantage over the EK method.
The technical disadvantage of the EK method has been pointed out in
\cite{cipriani2005}. The FC method can shorten the computation time
by several orders comparing with that with the EK method.

The paper is managed as follows. In next section we introduce the
two methods for studying energy diffusion in equilibrium one-dimensional
lattices. The rationale of the FC method is explained and the strategy
for applying this method is described. An example showing the qualitative
difference between the inner energy and the kick energy is reported.
Section III is contributed to investigating the energy diffusion in
stationary nonequilibrium lattices, i.e., in one-dimensional lattices
with stationary heat flux and stationary temperature gradients. This
section further reveals the diffusion behavior difference between
the inner energy and the kick energy. The last section is the conclusion
and the discussion of the paper.

\section{Energy diffusion in equilibrium lattices}

\begin{figure*}
\includegraphics{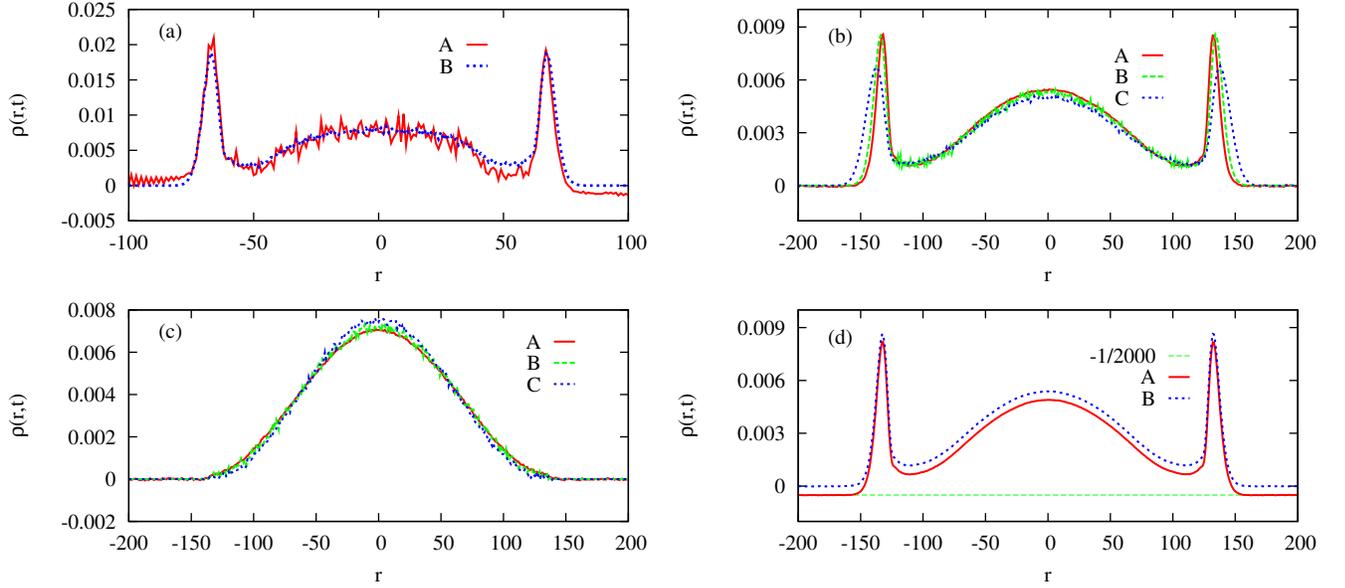}

\caption{(color online). The PDFs of energy diffusion at $T=0.5$. (a) The
PDFs at $t=50$ obtained by the EK method ($A$) and the improved
EK method ($B$) for the FPU-$\beta$ model. (b) The PDFs at $t=100$
obtained by the FC method($A$), and the EK method with $\Delta\widetilde{H}_{i}(0)=5E$
($B$) and $\Delta\widetilde{H}_{i}(0)=10E$ ($C$) for the FPU-$\beta$
model. (c) The PDFs at $t=200$ for the lattice $\phi^{4}$ model.
$A$, $B$ and $C$ are same as in (b). (d) The PDFs at $t=100$ for
the FPU-$\beta$ model obtained by the FC method using microcanonical
ensembles with eq.\ref{eq:cFC}($A$) and with eq.\ref{eq:mcFC} ($B$).
The line $C$ gives $\rho(r,t)=-1/N$ for reference.}

\label{fig1:EKvsFC}
\end{figure*}

A one-dimensional lattice is usually described by the Hamiltonian\[
H=\overset{N}{\underset{i=1}{\sum}}H_{i},\text{ }H_{i}=\frac{p_{i}^{2}}{2m}+U(x_{i})+V(x_{i}-x_{i-1}),\]
 where $U(x_{i})$ denotes the on-site potential and $V(x_{i+1}-x_{i})$
the interaction potential. In the present paper we study two types
of the lattice models, without or with on-site potentials. The illustrating
examples include the FPU-$\beta$ model with

\[
H_{i}=\frac{p_{i}^{2}}{2m}+\frac{1}{2}(x_{i}-x_{i-1})^{2}+\frac{1}{4}(x_{i}-x_{i-1})^{4}\]
 and the lattice $\phi^{4}$ model with

\[
H_{i}=\frac{p_{i}^{2}}{2m}+\frac{1}{2}(x_{i}-x_{i-1})^{2}+\frac{1}{4}x_{i}^{4}.\]
 Without loss of generality, we set mass $m=1$, and the Boltzmann
constant $k_{B}=1$ in following discussions. Fixed boundary condition
is applied for both models. We also employ occasionally the harmonic
model with $H_{i}=\frac{p_{i}^{2}}{2m}+\frac{1}{2}(x_{i}-x_{i-1})^{2}$,
the Toda model with $H_{i}=\frac{p_{i}^{2}}{2m}+e^{-(x_{i}-x_{i-1})}+(x_{i}-x_{i-1})-1$
and the quartic-FPU model with $H_{i}=\frac{p_{i}^{2}}{2m}+\frac{1}{4}(x_{i}-x_{i-1})^{4}$
for special purposes.

When adding heat baths to the two ends of a lattice one obtains a
canonical system. A canonical system is an open system and thus the
total energy of the system is not conservative. When heat the lattice
to a finite temperature and then take away the heat bath, one obtains
a microcanonical system. A microcanonical system is an isolated system
with total energy conserved. In both cases, after the stationary state
is achieved, every particle has identical temperature, $T=\left\langle p_{i}^{2}\right\rangle $,
and identical average energy, $E=\left\langle H_{i}\right\rangle $.

The problem to be investigated is: how does the energy $H_{i}$ or
the energy difference $\Delta H_{i}=H_{i}-E$ initially located at
the $i$th particle spread out over the lattice as a function of time?
To find the solution to this problem, the EK method adds a kick $\Delta\widetilde{H}_{i}(0)=\widetilde{H}_{i}(0)-H_{i}(0)$
to the $i$th particle at $t=0$, and calculate the part $\Delta\widetilde{H}_{j}(t)$
transported to the $j$th particle at time $t$, where $H_{i}(0)$
is the energy before the kick while $\widetilde{H}_{i}(0)$ represents
the energy after the kick. It is clear that without the external kick
the ensemble average $\left\langle \widetilde{H}_{j}(t)\right\rangle $
should be equal to the average energy $E$. Therefore, the ensemble
average $\left\langle \Delta\widetilde{H}_{j}(t)\right\rangle =\left\langle \widetilde{H}_{j}(t)\right\rangle -E$
can be explained as the average part of the kick energy transported
to the $j$th particle at time $t$. One can thereby define\begin{equation}
\rho(r,t)=\frac{\left\langle \Delta\widetilde{H}_{j}(t)\right\rangle }{\left\langle \Delta\widetilde{H}_{i}(0)\right\rangle }\label{eq:EK}\end{equation}
 as the PDF to describe the probability of the kick energy transported
to the position $r$ at time $t$, where $r=j-i$.

As an example, in Fig. \ref{fig1:EKvsFC}(a) we show the PDF obtained
by the EK method for the FPU-$\beta$ model with temperature $T=0.5$
and at $t=50$. The ensemble average is over $4.2\times10^{5}$ different
realizations. For each realization the lattice is developed to $t=50$,
starting by adding an external kick to the middle particle. For the
sake of simplicity we fix the amplitude of kicks at $\Delta\widetilde{H}_{i}(0)=5E$.
The figure indicates that the profile of the PDF is already distinguishable,
but the fluctuations are still large.

A different strategy can be applied to improve the convergence of
the PDF for the EK method \cite{delfini2007}. The idea is to copy
the system before the kick as a reference system. Then add the kick
to the middle particle of the original system and evolve the pair
of the systems simultaneously to compute the energy difference between
the two systems by $\Delta\widetilde{H}_{j}(t)=\widetilde{H}_{j}(t)-H_{j}(t)$,
where the first term in the r.h.s is the energy of the $j$th particle
of the kicked system while the second is that of the reference system.
It is clear that the difference represents exactly the kick energy
transported to the $j$th particle. By averaging the same amount of
realizations as in above we obtained the PDF and show it also in Fig.
\ref{fig1:EKvsFC}(a). One can find that the fluctuations are dramatically
smoothed. The PDFs obtained by the EK method hereafter are calculated
using this strategy.

The EK method is presented in the spirit of the linear response theory\cite{marconi2008},
which demands the kick energy small enough to keep a linear response.
However, practically, it can not be too small otherwise the fluctuations
of the ensemble average will be too large. In this situation, the
PDF will depend on the amplitude of the kick energy. In Fig. \ref{fig1:EKvsFC}(b)
we show the PDFs calculated with kicks of amplitude $\Delta\widetilde{H}_{i}(0)=5E$
and of amplitude $\Delta\widetilde{H}_{i}(0)=10E$. It is obvious
that the kick energy with $\Delta\widetilde{H}_{i}(0)=10E$ diffuses
faster than that with $\Delta\widetilde{H}_{i}(0)=5E$. The same effect
remains in the lattice $\phi^{4}$ model, as Fig. \ref{fig1:EKvsFC}(c)
shows. Nevertheless, in this model the high-energy kick diffuses slower
than that of the low-energy kick, which confirms the observation that
the higher of the energy, the stronger of the localization effect
in this model \cite{ponno2006}. Therefore, the EK method studies
indeed the diffusion process of the kick energy; it is not directly
explore the intrinsic diffusion behavior in the system.

\begin{figure}
\includegraphics{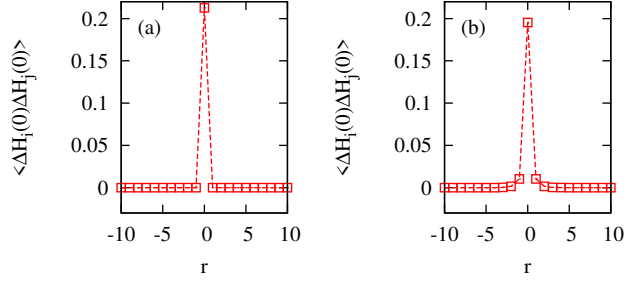}

\caption{(color online). The initial (spatial) correlation $\left\langle \Delta H_{i}(0)\Delta H_{j}(0)\right\rangle $
of the FPU-$\beta$ model (a) and the lattice $\phi^{4}$ model (b).}

\label{fig2:spt-Corr}
\end{figure}

The desirable way is to study directly the inner energy, i.e., the
energy initially localized at a particle or in a small region in the
lattice. The difficult is that one can not directly track this part
of energy because it can not be marked, as does in the case of particle
diffusion. Let $\Delta H_{i}(0)$ represents the energy fluctuation
initially located at the $i$th particle. Then if a portion of $\Delta H_{i}(0)$
is transported to the $j$th particle at time $t$, the energy fluctuation
$\Delta H_{j}(t)$ should have correlation with $\Delta H_{i}(0)$
physically. The idea of the FC method is to track the motion of the
inner energy by calculating the correlation of the energy fluctuations.
The first step of applying the FC method is to determine the part
of energy, denoted by $\Delta H(0)$, to be studied. We show how to
do this by examples. In Fig. \ref{fig2:spt-Corr}(a) and \ref{fig2:spt-Corr}(b),
we plot $\left\langle \Delta H_{i}(0)\Delta H_{j}(0)\right\rangle $
of the FPU-$\beta$ model and the lattice $\phi^{4}$ model respectively,
where $r=j-i$. Canonical systems with $N=501$ are employed for both
models. It can be seen that $\left\langle \Delta H_{i}(0)\Delta H_{j}(0)\right\rangle =0$
(within numerical errors) for $j\neq i$ and $\left\langle \Delta H_{i}(0)\Delta H_{j}(0)\right\rangle \neq0$
for $j=i$ for the FPU-$\beta$ model, which indicate that $\Delta H_{i}(0)$
has no initial correlation with the energy fluctuations of other particles.
In this case, we employ $\Delta H(0)=\Delta H_{i}(0)$ as the energy
to be studied. While in the case of the lattice $\phi^{4}$ model,
it has $\left\langle \Delta H_{i}(0)\Delta H_{j}(0)\right\rangle >0$
for several nearby particles, which implies that $\Delta H_{i}(0)$
is intrinsically correlated with a part of energy besides $\Delta H_{i}(0)$
itself. Because this part of energy can not be separated from $\Delta H_{i}(0)$,
we have to consider it as a whole as the objective $\Delta H(0)$
to be studied. We call $\Delta H(0)$ as the adherence energy of $\Delta H_{i}(0)$.

It is easy to understand why in certain systems $\Delta H_{i}(0)$
adheres a small part of energy while in other systems it does not.
In a lattice with finite temperature, at the same moment the fluctuations
of the kinetic energy of particles are independent because the velocity
of a particle at a moment puts no restriction on that of other particles.
For a class of lattices with only interaction potentials, such as
the FPU-$\beta$ model, the potential is determined by the relative
displacements. A configuration of $x_{i}-x_{i-1}$ which determines
the potential part of $H_{i}$ is completely independent of the configuration
of $x_{i+1}-x_{i}$ which determines the potential part of $H_{i+1}$.
As a result, $H_{i}$ and $H_{i+1}$ are independent with each other
and thus $\left\langle \Delta H_{i}(0)\Delta H_{j}(0)\right\rangle =0$
for $j\neq i$. For another class of lattices with on-site potentials,
such as the lattice $\phi^{4}$ model, the situation is different.
When the $i$th particle moves to the left/right to its equilibrium
position, the neighbor particles will tends to follow it because of
the interaction potential between them. This effect results in a positive
correlation, i.e., the bigger the on-site potential $x_{i}^{4}/4$,
the bigger the $x_{i\pm1}^{4}/4$, and \emph{vice versa}.

We then define\begin{equation}
\rho(r,t)=\frac{\left\langle \Delta H_{i}(0)\Delta H_{j}^{\prime}(t)\right\rangle }{\left\langle \Delta H_{i}(0)\Delta H(0)\right\rangle }\label{eq:dFC}\end{equation}
 as the PDF to describe the motion of $\Delta H(0)$. The reason is
as follows. First, $\Delta H(0)$ never disappears before it spreads
out of the lattice because of the energy conservation, i.e., it is
a conserved quantity. Thus, one has $\sum_{j}\Delta H_{i}(0)\Delta H_{j}^{\prime}(t)=\Delta H_{i}(0)\Delta H(0)$,
which gives $\int\rho(r,t)dx=1$. Second, $\Delta H_{i}(0)$ and $\Delta H(0)$
are positively correlated, and $\Delta H_{j}^{\prime}(t)$ and $\Delta H(0)$
should also be positively correlated since the former is a part of
the latter, one obtains $\rho(r,t)\geqslant0$. Non-negativity and
normalization conditions are basic requirements to be a PDF. Finally,
it is clear that $\rho(r,t)$ is proportional to $\Delta H_{j}^{\prime}(t)/\Delta H(0)$.
Therefore, $\rho(r,t)$ can be considered as the probability of finding
$\Delta H(0)$ at the position $r$ at time $t$, or equally, it represents
the rate of $\Delta H(0)$ to be transported to $r$ at time $t$.

The problem is how to calculate $\left\langle \Delta H_{i}(0)\Delta H_{j}^{\prime}(t)\right\rangle $.
At time $t$ we divide the energy fluctuation at the $j$th particle
into two parts, i.e., $\Delta H_{j}(t)=\Delta H_{j}^{\prime}(t)+\Delta H_{j}^{\prime\prime}(t)$,
where $\Delta H_{j}^{\prime}(t)$ represents the part of $\Delta H(0)$
being transported to the $j$th particle, and $\Delta H_{j}^{\prime\prime}(t)$
comes from other sources. For a canonical system, $\Delta H_{j}^{\prime\prime}(t)$
has no correlation with $\Delta H_{i}(0)$ since it comes from \{other
parts of the system or\} the heat baths. In this case, it should has
$\left\langle \Delta H_{i}(0)\Delta H_{j}^{\prime\prime}(t)\right\rangle =0$.
As a result, we can calculate $\left\langle \Delta H(0)\Delta H_{j}^{\prime}(t)\right\rangle $
by $\left\langle \Delta H(0)\Delta H_{j}^{\prime}(t)\right\rangle =\left\langle \Delta H(0)\Delta H_{j}(t)\right\rangle $
and $\left\langle \Delta H_{i}(0)\Delta H(0)\right\rangle $ by $\left\langle \Delta H_{i}(0)\Delta H(0)\right\rangle =\sum_{j}\left\langle \Delta H_{i}(0)\Delta H_{j}(0)\right\rangle $.
In other word, in studying canonical systems one can apply\begin{equation}
\rho(r,t)=\frac{\left\langle \Delta H_{i}(0)\Delta H_{j}(t)\right\rangle }{\left\langle \Delta H_{i}(0)\Delta H(0)\right\rangle }\label{eq:cFC}\end{equation}
 to calculate the PDF of $\Delta H(0)$.

For a microcanonical system, however, the condition $\left\langle \Delta H_{i}(0)\Delta H_{j}^{\prime\prime}(t)\right\rangle =0$
fails because of the conservation of the total energy of the system.
The conservation of the total energy implies $\sum_{j}\Delta H_{j}(t)=0$,
which can be rewritten as $\sum_{j}[\Delta H_{j}^{\prime}(t)+\Delta H_{j}^{\prime\prime}(t)]=\Delta H(0)+\sum_{j}\Delta H_{j}^{\prime\prime}(t)=0$.
Multiplying $\Delta H_{i}(0)$ to the equation one obtains $\sum_{j}\Delta H_{i}(0)\Delta H_{j}^{\prime\prime}(0)=-\Delta H_{i}(0)\Delta H(0)$.
By calculating the ensemble average it appears as $\sum_{j}\left\langle \Delta H_{i}(0)\Delta H_{j}^{\prime\prime}(t)\right\rangle =-\left\langle \Delta H_{i}(0)\Delta H(0)\right\rangle $.
Suppose that the lattice includes $N$ particles. Because the nonvanishing
correlation $\left\langle \Delta H_{i}(0)\Delta H_{j}^{\prime\prime}(t)\right\rangle $
is resulted from the conservation of the total energy, it is an intrinsic
correlation of the system and should be independent of time. It is
reasonable to assume that $\left\langle \Delta H_{i}(0)\Delta H_{j}^{\prime\prime}(t)\right\rangle $
should have identical value for each particle. One then derives

\[
\left\langle \Delta H_{i}(0)\Delta H_{j}^{\prime\prime}(t)\right\rangle =-\frac{1}{N}\left\langle \Delta H_{i}(0)\Delta H(0)\right\rangle .\]
 Notice that $\Delta H_{j}^{\prime}(t)=\Delta H_{j}(t)-\Delta H_{j}^{\prime\prime}(t)$,
the PDF of the energy diffusion in a microcanonical system appears
as\begin{equation}
\rho(r,t)=\frac{\left\langle \Delta H_{i}(0)\Delta H_{j}(t)\right\rangle }{\left\langle \Delta H_{i}(0)\Delta H(0)\right\rangle }+\frac{1}{N}.\label{eq:mcFC}\end{equation}

\begin{figure}
\includegraphics{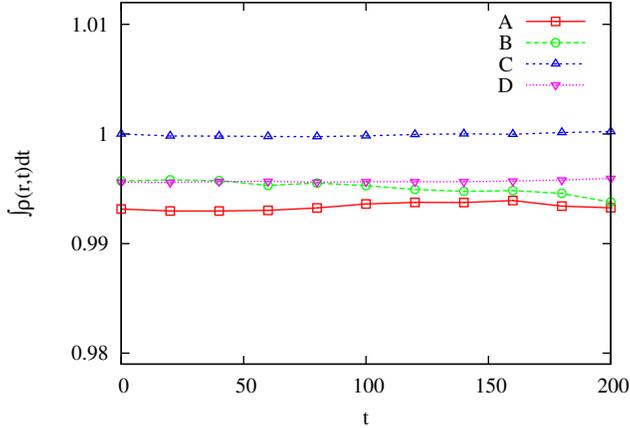}

\caption{(color online). $\int\rho(r,t)dr$ \emph{versus} time $t$ for the
FPU-$\beta$ model obtained with canonical ensemble($A$) and microcanonical
ensemble ($B$), and the lattice $\phi^{4}$ model with canonical
ensemble($C$) and microcanonical ensemble ($D$)}

.

\label{fig3:norm}
\end{figure}

These two equations \ref{eq:cFC} and \ref{eq:mcFC} trustily represent
the definition equation \ref{eq:dFC} since one can check that the
$\rho(r,t)$ obtained by them satisfy the conditions $\int\rho(r,t)dx=1$
and $\rho(r,t)\geqslant0$. In Fig. \ref{fig1:EKvsFC}(b) and \ref{fig1:EKvsFC}(c),
we show the $\rho(r,t)$ calculated by the FC method using canonical
ensembles. One can see that for either the FPU$-\beta$ model or the
lattice $\phi^{4}$ model the condition $\rho(r,t)\geqslant0$ is
satisfied within numerical errors. In Fig. \ref{fig3:norm}, we show
$\int\rho(r,t)dx$ as a function of time for the two models at $T=0.5$,
which indicates that $\int\rho(r,t)dx=1$ is also satisfied with high
precision.

To obtain an isolated system with given temperature, we heat the model
to the stationary state with the expected temperature and then take
away the heat bath. Applying such a microcanonical system to calculate
the PDF, one should employ the eq. \ref{eq:mcFC} instead of the eq.
\ref{eq:cFC}. In Fig. \ref{fig1:EKvsFC}(d), the dotted line shows
the $\rho(r,t)$ calculated by eq. \ref{eq:mcFC} while the solid
line shows that of calculated by eq. \ref{eq:cFC} in the case of
the isolated FPU$-\beta$ model. It can be seen that the condition
$\rho(r,t)\geqslant0$ is well satisfied in the former case while
fails in the latter case. The condition $\int\rho(r,t)dx=1$ is also
satisfied within numerical errors as shown in Fig. \ref{fig3:norm}
in the former case, while it can be checked that $\int\rho(r,t)dx=0$
in the latter case. Moreover, we have checked that the PDFs obtained
with the canonical ensemble and the microcanonical ensemble are identical
with each other at the same temperature within numerical errors.

From Fig. \ref{fig1:EKvsFC}(b) and \ref{fig1:EKvsFC}(c) one can
realize that a PDF obtained by the EK method will approach the corresponding
PDF obtained by the FC method when the kick energy $\Delta\widetilde{H}_{i}(0)$
is small enough. However, the smaller the kick, the slower the convergence
of the PDF. As can be seen in Fig. \ref{fig1:EKvsFC}(b), with the
same amount of the realizations, the fluctuations of the PDF obtained
with the kicks $\Delta\widetilde{H}_{i}(0)=5E$ is much larger than
that of the PDF obtained with $\Delta\widetilde{H}_{i}(0)=10E$. Therefore,
decreasing $\Delta\widetilde{H}_{i}(0)$ needs to dramatically increase
the amount of the realizations to achieve a reliable ensemble average.

We now describe the technical details in applying the two methods.
Let us take the FPU-$\beta$ model as an example. For this model,
as shown in \cite{zhao2006}, the two soliton-like packets on the
PDF move at a constant velocity, which is supersonic and depends on
the temperature of the system. With the dimensionless unit, the soliton-like
packets move with $v>1$. In this case, supposing that the diffusion
starts at the middle particle of the lattice and one wants to study
the diffusion process up to a time length $t=t_{c}$, the lattice
must have a size $N>2t_{c}$ to keep the energy diffusing within the
lattice during this period.

To apply the EK method, one may employ a lattice with length $N+1$
and evolve the system with a sufficiently long time to relax it to
a stationary state. Then copy a reference system before adding a kick
$\Delta\widetilde{H}_{i}(0)$ to the middle particle of the lattice.
And evolve both the reference system and the kicked system for a time
$t_{c}$ to obtain a set of realization $\Delta\widetilde{H}_{j}(t)$.
This step, i.e., kicking and evolving the two systems, is repeated
again and again to obtain the ensemble average $\left\langle \Delta\widetilde{H}_{j}(t)\right\rangle $.
Each round of repeat contributes one realization to the ensemble.
In this way, evolving a couple of systems for a period $t$ one obtains
$t/t_{c}$ realizations to the ensemble.

\begin{figure}
\includegraphics{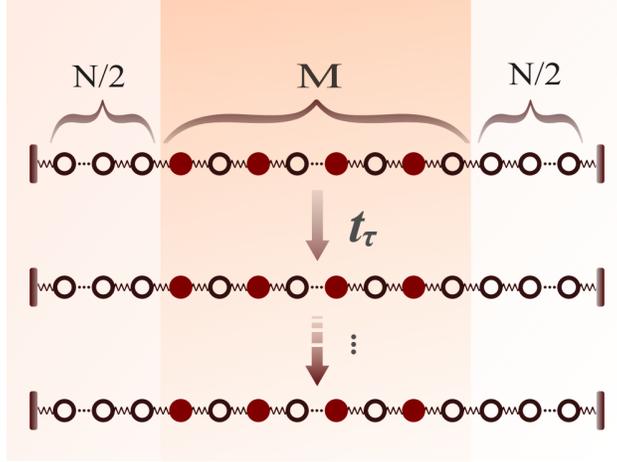}

\caption{(color online). Illustration of the strategy of the FC method for
calculating the ensemble average. Each of the black particles is applied
as independent source where the diffusion is starting at.}

\label{fig4:ensemble}
\end{figure}

To apply the FC method, we apply a lattice with a size $N/2+M+N/2$,
and select $M/\xi_{\tau}$ particles in the middle segment as the
sources that the diffusion starts, as illustrated in Fig. \ref{fig4:ensemble}.
Here $\xi_{\tau}$ is a constant which should be sufficiently bigger
than the spatial correlation length between particles. For the FPU-$\beta$
model, one can set $\xi_{\tau}=1$ since Fig. \ref{fig2:spt-Corr}(a)
has indicated that there is no spatial correlations between different
particles. For the lattice $\phi^{4}$ model, Fig. \ref{fig2:spt-Corr}(c)
has shown that the correlation of a particle with its next nearest
neighbors has closely approached zero and one can set $\xi_{\tau}=3$.
In this way, each selected particle can be considered as an independent
source and contributes realizations to the ensemble average independently.
Based on the same principle, for each of the selected particles, one
can further apply a set of continuous records, $\Delta H_{i}(0)$,
$\Delta H_{i}(t_{\tau})$, $\Delta H_{i}(2t_{\tau})$, $\Delta H_{i}(3t_{\tau})$,$\cdots$,
as the independent sources, where $t_{\tau}$ is a constant which
should be sufficiently bigger than the autocorrelation length of the
particle. For the FPU-$\beta$ and the $\phi^{4}$ models, it can
be easily checked that one can apply $t_{\tau}=2$ and $t_{\tau}=3$
respectively. In this way, each solid circle in Fig. \ref{fig4:ensemble}
is applied as an independent source of energy diffusion, and contributes
a realization to the ensemble.

In this way, by evolving the lattice with a time $t$ one can obtain
$M(t-t_{c})/\xi_{\tau}t_{\tau}$ realizations to the ensemble, which
is about $Mt_{c}/\xi_{\tau}t_{\tau}$ times of those obtained in the
case of the EK method if $t\gg t_{c}$.

In the FPU-$\beta$ model with temperature $T=0.5$, for example,
the soliton-like packets on the PDF move with a velocity $v\sim1.3$.
To investigate the diffusion process extended to $t_{c}=1000$, one
can employ a lattice with $N=3001$ when applying the EK method and
apply the middle particle as the source that the energy starts to
diffuse. While applying the FC method, one may employ a lattice with
$4000$ particles and selects the source particles from the middle
segment with $M=1000$. In both cases, the energy diffusion remains
within the lattices during the period $t_{c}$, since within this
time the soliton-like packets can spread over about $1300$ particles.
Evolving the two lattices with the same time $t$, the realizations
to the ensemble obtained by the FC method is about $Mt_{c}/\xi_{\tau}t_{\tau}\sim5\times10^{5}$
times of the realizations obtained by the EK method.

\begin{figure}
\includegraphics{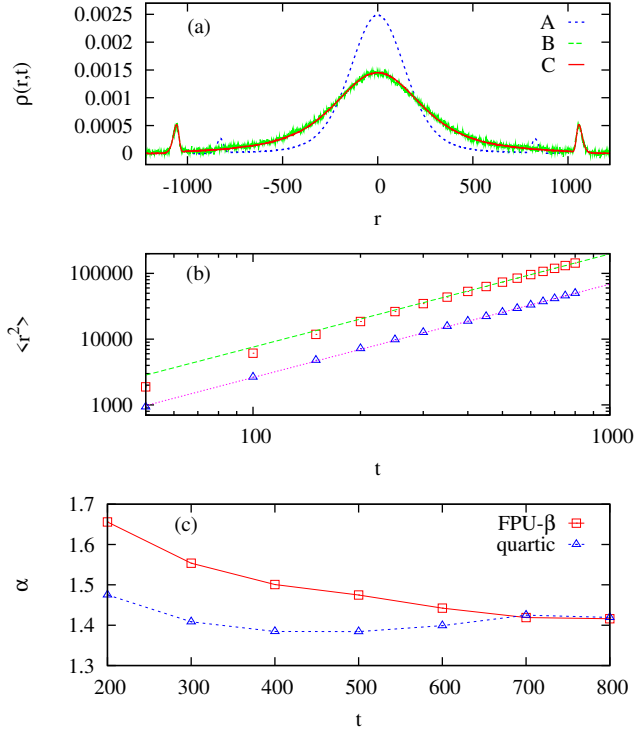}

\caption{(color online). The finite-time effect of the diffusion exponent $\alpha$.
(a) The PDFs at $t=800$: $A$, for the quartic-FPU model averaged
over $6\times10^{11}$ realizations; $B$ and $C$, for the FPU-$\beta$
model averaged over $3\times10^{9}$ and and $3\times10^{11}$ realizations
respectively. (b) The log-log scale plot of $\left\langle r^{2}(t)\right\rangle $
against $t$ for the two models. (c)The time-dependent effect of $\alpha$
for the two models. In both (b) and (c), the squares represent the
results of the FPU-$\beta$ model and the triangles are for the quartic-FPU
model.}

\label{fig5:t800}
\end{figure}

Figure \ref{fig5:t800} (a) shows the PDFs of the FPU-$\beta$ model
at $t_{c}=800$ calculated by averages over $3\times10^{9}$ and $3\times10^{11}$
realizations respectively. It can be seen that the fluctuations of
the PDF are suppressed to a reasonable level in the latter case while
is still remarkable in the former case. Even using the FC method,
to obtain $3\times10^{11}$ realizations for such a long time diffusion
processes is still a hard task for serious computation, to say nothing
of the EK method. Fortunately, we achieved the goal by parallel computations
using $28$ CPUs with half a month. If one wants to obtain the same
amount of realizations by the EK method, he must use about $5\times10^{6}$
CPUs with the same computation time.

Achieving a sufficiently long diffusion time is necessary to avoid
the finite-time effect. The finite-time effect can be ignored for
the lattice $\phi^{4}$ model, in which the exponent $\alpha$ tends
to $\alpha=1$ within $t_{c}<100$. For other types of lattices with
anomalous energy diffusion, such as the FPU-$\beta$ model and the
quartic-FPU model, the finite-time effect is remarkable. The quartic-FPU
model represents the unharmonic (or high-temperature) limit of the
FPU-$\beta$ model. As pointed in \cite{zhao2005}, the qualitative
statistical property of this model is temperature-independent because
of the scaling behavior. In other words, systems with different temperatures
can be scaled together by a proper scaling transformation.

As shown in Fig. \ref{fig5:t800}(a), the PDFs of the energy diffusion
for the two models are qualitatively similar with each other, though
the soliton-like packets on the PDF of the quartic-FPU model are obviously
smaller than that of the FPU-$\beta$ model. In Fig. \ref{fig5:t800}(b),
we plot the $\left\langle r^{2}(t)\right\rangle $ as a function of
time with the log-log scale for the two models. Roughly, it seems
that both models show a power-law relationship between $\left\langle r^{2}(t)\right\rangle $
and $t$. To explore the finite-time effect, we plot $\alpha$ by
sectional fitting, as Fig. 6(c) shown. In the plot, each point of
$\alpha$ is obtained by fitting 4 data points, i.e., it represents
the result in a time interval $\Delta t=200$ correspondingly. One
can see that $\alpha$ changes with time initially and converges gradually
to a constant, and the finite-time-dependent behavior of $\alpha$
for the two models have different features. It can be seen also that
$\alpha$ approaches $1.41$ at $t_{c}>600$ for both models. This
result tends to support the relationship equation of $\alpha$ and
$\beta$ presented in Ref. \cite{denisov2003}. However, we still
can not sure whether the diffusion exponents have converged exactly
at $t_{c}=800$. To check it one needs to extend the computation to
a longer diffusion time, which already exceeds the capability of our
computer system. At least, this fact reminds one to be careful in
checking the relationship between $\alpha$ and $\beta$ by numerical
calculations. Instead of studying the diffusion process with the time
scale of $t_{c}\sim100$ as done by some previous researchers, a diffusion
time $t_{c}>700$ at least is needed to avoid the finite-time effect.

\begin{figure}
\includegraphics{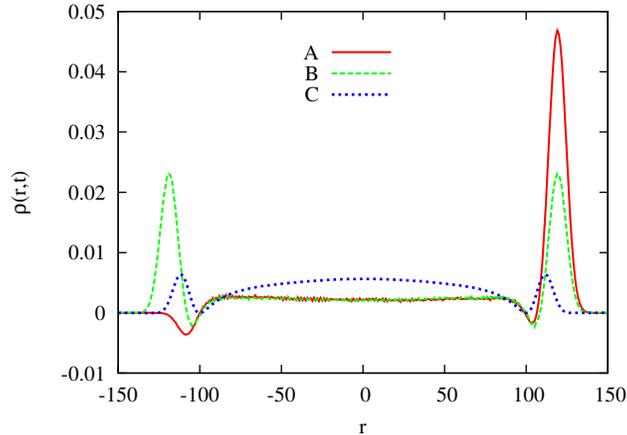}

\caption{(color online). The PDFs of the Toda model at $T=0.5$ and $t=100$:
$A$, by the EK method with positive kicks; $B$, by the EK method
using the same number of positive and negative kicks; $C$ by the
FC method.}

\label{fig6:Toda}
\end{figure}

According to the above results, it seems that the PDFs obtained by
the two methods are always symmetric and have only quantitative differences.
The symmetric PDFs obtained by the FC method represent the intrinsic
behavior of the inner energy. Any part of energy must moves in symmetric
way statistically, otherwise the equilibrium can not be maintained.
However, the symmetric PDFs obtained by the EK method are particular
characteristic of a class of special lattices. By reconsidering the
models discussed above, one can find that they have a common feature,
i.e., the interaction potential in these models are symmetric. As
a consequence, the kick energy added on a particle, either with positive
or negative momentum, will be equally transported to the opposite
directions. In these lattices, the PDFs obtained by the EK method
are always symmetric. On the contrary, for lattices with asymmetric
interaction potentials, the PDF obtained by the EK method depends
on the way of the kicks. A typical model with asymmetric interaction
potential is the Toda model. Figure \ref{fig6:Toda} shows the PDFs
computed by the FC and the EK methods respectively for this model
in equilibrium with $T=0.5$. Physically, the inner energy must have
equal probability to move to both sides to maintain the equilibrium,
the PDF should be symmetric. The figure indicates that the FC method
explores this feature correctly. To applying the EK method, we add
the kicks in two ways. One is to always add kicks with positive momenta.
Notice the asymmetric feature of the interaction potential in this
model, a kick with positive momentum will transport more energy to
the r.h.s than to the l.h.s. As a result, the PDF of the kick energy
should be asymmetric, which is confirmed by the figure. Another way
is to add positive and negative kicks with equal probability. In this
case, one can obtain a symmetric PDF as the figure shows. Thus, the
PDF of the kick energy depends on the way of the kicks. Furthermore,
the figure explores another serious problem of the EK method. While
the condition $\rho(r,t)\geqslant0$ well-satisfied for the PDF obtained
by the FC method, it fails for the EK method in certain intervals,
and thus fails to be a probabilistic density function. Therefore,
for this model, the EK method is indeed unsuitable for studying the
diffusion process.

\section{Energy diffusion in nonequilibrium lattices}

\begin{figure*}
\includegraphics{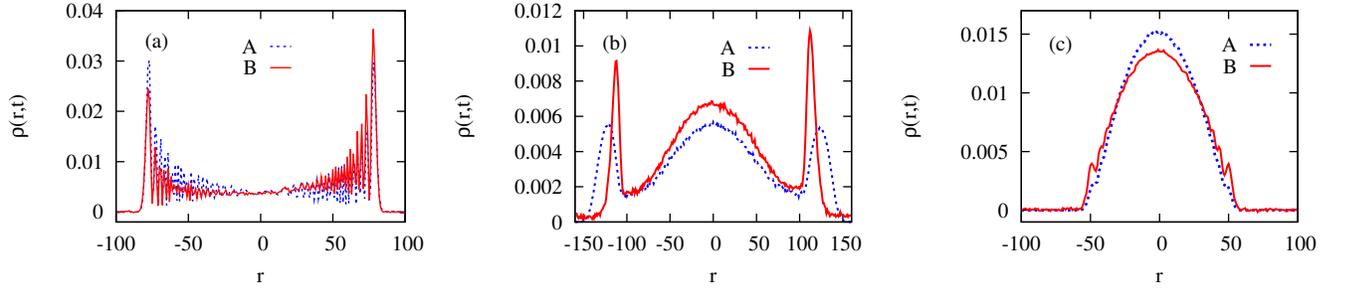}

\caption{(color online). The PDFs at $t=80$ calculated in the case of stationary
noequilibrium systems by the EK method ($A$) and the FC method ($B$).
(a) The harmonic model. (b) The FPU-$\beta$ model. (c) The lattice
$\phi^{4}$ model. In each model, the lattice size is $N=501$ and
the temperatures of the heat baths are fixed at $T_{+}=1$ and $T_{-}=0.5$.}

\label{fig7:nonequilibrium}
\end{figure*}

Studying the diffusion in stationary nonequilibrium lattices provides
us more clear examples to reveal the different diffusion behavior
of the inner energy and the kick energy. When coupled with two heat
baths with different temperatures, a lattice will approach a nonequilibrium
stationary state after a relaxation process. A constant heat flux
is then established along the lattice. In such a nonequilibrium lattice,
the temperature as well as the average energy can only be defined
locally, i.e., $T_{i}=\left\langle p_{i}^{2}\right\rangle $, and
$E_{i}=\left\langle H_{i}\right\rangle $. The $\Delta H_{i}(0)=H_{i}(0)-E_{i}$
depends also on the position of the particles, and, therefore, the
energy diffusion starting at different particles may not be statistically
identical. In this situation, we study the diffusion starting at the
middle particle of a lattice as the illustrating example. Consequently,
in a nonequilibrium lattice, one can no longer employ a segment with
$M$ particles to calculate the ensemble average, as doing in the
case of equilibrium systems. However, the FC method still has the
technique advantage in calculating the ensemble average since along
the time axis the strategy described in Fig. \ref{fig4:ensemble}
is still available.

In Fig. \ref{fig7:nonequilibrium}, we show the PDFs obtained by the
EK method (dotted lines) and the FC method (solid lines) for the harmonic
model, the FPU-$\beta$ model and the lattice $\phi^{4}$ model. The
temperature of the heat bath in the l.h.s is $T_{+}=1$ and is $T_{-}=0.5$
in the r.h.s. Each lattice includes $N=501$ particles and the diffusion
time is $t=80$. It can be seen that except of the $\phi^{4}$ model,
the PDFs obtained by the EK method are symmetric while those by the
FC method are asymmetric. In more detail, it can be checked that $\sigma\simeq1$
for the PDFs obtained by the EK method. For the PDFs obtained by the
FC method, it has $\sigma\simeq1.4$ for the harmonic model, $\sigma\simeq1.1$
for the FPU-$\beta$ model, and $\sigma\simeq1$ for the lattice $\phi^{4}$
model.

The PDFs with $\sigma\simeq1$ obtained by the EK method represent
the diffusion behavior of the kick energy. As has been pointed out
in the end of last section, for lattices with symmetric interaction
potentials the kick energy is transported to the opposite directions
equally. Either in equilibrium or in nonequilibrium, the environment
around the middle particle is similar locally, and thus the kick energy
is divided into two equal parts for this type of lattice. The two
parts of energy are transported towards the opposite directions hereafter.

The FC method correctly explores the diffusion behavior of the inner
energy. The harmonic model is the only one been solved rigidly for
the heat conduction problem \cite{rieder1967}. The mechanism of
the heat transport in this model is clear, i.e., the energy delivered
by one heat bath will be transported to the opposite side directly
since there is no interaction among normal modes (phonons) in this
model. As a result, the energy fluctuation $\Delta H_{i}(0)$ at the
middle particle can be exactly divided into two parts $\Delta H_{i}(0)=\Delta H_{i}^{+}(0)+\Delta H_{i}^{-}(0)$.
The part $\Delta H_{i}^{+}(0)$ comes from the l.h.s heat bath and
will move to the r.h.s, and the part $\Delta H_{i}^{-}(0)$ comes
from the r.h.s heat bath and will move to the l.h.s. In the case of
$T_{+}=T_{-}$, it has $\Delta H_{i}^{+}(0)=\Delta H_{i}^{-}(0)$
and in a later time the amount of the energy $\Delta H_{i}(0)$ transported
to both sides of the middle particle will be the same. While in the
case of $T_{+}>T_{-}$, it has $\Delta H_{i}^{+}(0)>\Delta H_{i}^{-}(0)$
and in a later time the energy moves to the left will be more than
that moves to the right. This is the expected diffusion property of
the inner energy in this lattice. We define a parameter $\sigma$
as \[
\sigma\equiv\frac{\int_{0}^{\infty}\rho(r,t)dx}{\int_{-\infty}^{0}\rho(r,t)dx}\]
 to measure the rate of the energy spreading to the two sides of the
middle particle. When applying the FC method, it is easy to obtain
$\sigma=\frac{\left\langle \Delta H_{i}(0)\Delta H_{i}^{+}(0)\right\rangle }{\left\langle \Delta H_{i}(0)\Delta H_{i}^{-}(0)\right\rangle }$.
One thus derives $\sigma=1$ in equilibrium lattices and $\sigma>1$
in nonequilibrium lattices, since $\Delta H_{i}^{+}(0)=\Delta H_{i}^{-}(0)$
in the former and $\Delta H_{i}^{+}(0)>\Delta H_{i}^{-}(0)$ in the
latter cases.

In the case of the FPU-$\beta$ model, as pointed out in \cite{zhao2006},
the soliton-like packets keep the initial memory of the direction
wherever they move. Equivalently, in this model the energy in the
soliton-like packets keeps its initial memory of moving direction.
As a result, the energy fluctuation $\Delta H_{i}(0)$ at the middle
particle can be divided into three parts, $\Delta H_{i}(0)=\Delta H_{i}^{+}(0)+\Delta H_{i}^{-}(0)+\Delta H_{i}^{0}(0).$
Here $\Delta H_{i}^{+}(0)$ and$\ \Delta H_{i}^{-}(0)$ come from
the high- and low-temperature heat baths and still keep the memory
of their moving directions; they will travel to the opposite sides.
The part $\Delta H_{i}^{0}(0)$ represents the no-memory part which
has equal probability to move to either direction. Because of $\Delta H_{i}^{+}(0)>\Delta H_{i}^{-}(0)$,
one can still predict $\sigma=\frac{\left\langle \Delta H_{i}(0)\Delta H_{i}^{+}(0)\right\rangle +\left\langle \Delta H_{i}(0)\Delta H_{i}^{0}(0)\right\rangle }{\left\langle \Delta H_{i}(0)\Delta H_{i}^{-}(0)\right\rangle +\left\langle \Delta H_{i}(0)\Delta H_{i}^{0}(0)\right\rangle }>1$.

For the lattice $\phi^{4}$ model, the Gaussian PDF as well as the
derived property of $\left\langle r^{2}(t)\right\rangle \sim t^{0}$
indicates that the energy diffuses in this system normally. Normal
diffusion implies a no-memory random walk. The energy starting at
any point will loss totally the initial memory of its moving direction.
Thus, $\Delta H_{i}(0)$ at the middle particle has no memory of the
heat baths and will diffuse to either side with equal probability.
In this case, one should expect $\sigma=1$ for this model.

\section{Summary and discussion}

In summary, the FC method reveals the intrinsic diffusion behavior
of the energy in the system, while the EK method displays the diffusion
behavior of the kick energy. The inner energy and the kick energy
may carry different information, and they may diffuse in different
ways.

In lattices with normal energy diffusion, such as in the lattice $\phi^{4}$
model, the energy will totally lose the initial memory. In these lattices,
the behavior of the kick and the inner energy appear qualitatively
similar because both of them will lose the initial information totally.
The PDFs obtained by the two methods appear as Gaussian functions,
either in equilibrium or in nonequilibrium. However, the results of
the two methods are quantitatively different. The PDFs obtained by
the EK method depend on the amplitudes of the kick energy, only when
the kicks are weak enough can it approach the results of the FC method.

In lattices with anomalous energy diffusion, the energy may keep part
of the initial memory. In equilibrium case, the memory informtion
about the moving directions included in the inner energy is symmetric,
thus the PDFs obtained by the FC method exhibits such symmetry. When
the interaction potentials in lattices, such as the harmonic model
or the FPU-$\beta$ model, are symmetric, the kick energy is thus
divided into two equal parts, one moves to the right and the other
to the left. As a result, the PDFs obtained by the EK method are also
symmetric, although they may be quantatively different from those
obtained by the FC method. When the interaction potential is asymmetric,
such as in the Toda model, however, the kick energy may diffuse asymmetrically,
depending on the way how the kicks are added on. Because of the asymmetric
potential, a kick to the right, for instance, will transport more
energy to the r.h.s than to the l.h.s., and vice versa. In this kind
of lattices, if one adds positive and negative kicks to the particle
with equal probability, the PDFs calculated by the EK method are symmetric,
otherwise the PDFs must appear asymmetrically.

In nonequilibrium case, the heat baths added on the two ends of a
lattice have different temperatures. As a result, the parts of energy
keeping the memory of the corresponding heat bathes arrived at the
middle particle are different. The PDFs obtained by the FC method
are asymmetric and reveal this intrinsic feature of inner enrgy, while
the EK method fails to do so because the kick energy keeps no information
of memory of the heat bathes.

Besides the solid theoretical basis, the FC method has remarkable
technique advantage over the EK method. Because the FC method uses
the energy fluctuations of particles in the system, and any independent
energy fluctuation can be treated as $\Delta H_{i}(0)$ to calculate
the ensemble average, thus one can obtain a large set of independent
realizations to the ensemble by one round of evolution of the system,
as illustrated in Fig. \ref{fig4:ensemble}. This strategy can reduce
the computation time dramatically. The adventage becomes quite remarkable
when one studies the long-time diffusion behavior. Comparing with
the EK method, the time to evolve the system is shortened by at least
five orders when one studies a diffusion process extended to $t_{c}=1000$.
With this strategy, we calculated the diffusion exponents $\alpha$
of the FPU-$\beta$ model and the quartic-FPU model upto $t=800$,
and observed that the size-effect on $\alpha$ obviously exist till
at least $t=600$. This fact indicates that the $\alpha$ calculated
in some previous works by the EK method with the diffusion time upto
$t\sim100$ is questionable.

To apply the FC method, we introduced the formulae for calculating
the PDF of energy diffusion available for canonical and micro-canonical
ensembles respectively. In the formulae, the conserved quantity being
investigated is $\Delta H(0)$ instead of $\Delta H_{i}(0)$. This
is because that in certain models the energy $\Delta H_{i}(0)$ adheres
constantly a part of energy. To guarantee the normalization of $\rho(r,t)$,
we have to consider $\Delta H(0)$ as the energy to be studied. Therefore,
the first step to apply the FC method is to detect the energy $\Delta H(0)$
adhered to $\Delta H_{i}(0)$ by checking the correlation $\left\langle \Delta H_{i}(0)\Delta H_{j}(0)\right\rangle $.
The energy $\Delta H(0)$ needs not to be calculated explicitly, but
we demand it locating in a small region and positively correlated
to $\Delta H_{i}(0)$. When these conditions are fulfilled, $\Delta H(0)$
is available for probabilistic descriptions and the $\rho(r,t)$ describes
the probability of finding $\Delta H(0)$ at position $r$ and time
$t$.

The definition of the PDF for canonical system presented in this paper
is slightly different from that presented in the Ref. \cite{zhao2006},
in which $\Delta H_{i}(0)$ is directly applied as the conserved quantity
to be studied. This treatment results no problem for certain lattices
such as the FPU-$\beta$ model. In such a lattice, it has $\Delta H(0)=\Delta H_{i}(0)$
because of $\left\langle \Delta H_{i}(0)\Delta H_{j}(0)\right\rangle =0$
for $i\neq j$. For other systems, such as the lattice $\phi^{4}$
model, $\Delta H(0)$ is slightly bigger than $\Delta H_{i}(0)$.
As a result, the $\rho(r,t)$ obtained can not be normalized exactly.
With higher precision, we have checked that $\int\rho(r,t)dx\sim1.07$
for the lattice $\phi^{4}$ model at $T=0.5$ when applying the previous
definition. Because $\int\rho(r,t)dx$ is still time-independent,
it will not alter the value of the exponent $\alpha$. However, as
a PDF, normalization is an expected feature. It is thus better to
consider $\Delta H(0)$ as the conservated quantity to be studied.

\begin{acknowledgments}
This work is supported by the National Natural Science Foundation
of China under Grant No. $10775115$, National Basic Research Program
of China ($973$ Program) ($2007CB814800$), and NCETXMU of Xiamen
University.
\end{acknowledgments}


\begin{thebibliography}{33}
\bibitem{metzler2000}
R. Metzler and J. Klafter, Phys. Rep. \textbf{339}, 1 (2000).
\bibitem{zaslavsky2002}
G.M. Zaslavsky, Phys. Rep. \textbf{371}, 461 (2002).
\bibitem{perrin1909}
J. Perrin, Ann. Chim. Phys. \textbf{18}, 5 (1909).
\bibitem{marrero1972}
T. Marrero and E. Mason, J. Phys. Chem. Ref. Data. \textbf{1}, 3 (1972).
\bibitem{livi2003}
R. Livi and S. Lepri, Nature \textbf{421}, 327 (2003).
\bibitem{lepri2003}
S. Lepri, R. Livi and A. Politi, Phys. Rep. \textbf{377}, 1 (2003).
\bibitem{rieder1967}
Z. Rieder, J.L. Lebowitz  and E. Lieb, J. Math. Phys. \textbf{8}, 1073 (1967).
\bibitem{lepri1997}
S. Lepri,  R. Livi and A. Politi, Phys. Rev. Lett. \textbf{78}, 1896 (1997).
\bibitem{hatano1999}
T. Hatano,  Phys. Rev. E \textbf{59}, R1 (1999).
\bibitem{lepri2000}
S. Lepri, Eur. Phys. J. B. \textbf{18}, 441 (2000).
\bibitem{dhar2001}
A. Dhar, Phys. Rev. Lett. \textbf{86}, 5882 (2001).
\bibitem{casati1984}
G. Casati,  J. Ford,  F. Vivaldi and  W.M. Visscher,  Phys. Rev. Lett. \textbf{52},  1861 (1984).
\bibitem{hu1998}
B. Hu, B. Li and H. Zhao, Phys. Rev. E \textbf{57}, 2992 (1998).
\bibitem{hu2000}
B. Hu, B. Li and H. Zhao, Phys. Rev. E \textbf{61},  3828 (2000).
\bibitem{Giardina2000}
C. Giardin{\'a},  R. Livi,  A. Politi and  M. Vassalli,  Phys. Rev. Lett. \textbf{84}, 2144 (2000).
\bibitem{Gendelman2000}
O.V. Gendelman  and A.V. Savin,  Phys. Rev. Lett. \textbf{84},  2381 (2000).
\bibitem{libw2003}
B. Li and  J. Wang,  Phys. Rev. Lett. \textbf{91},  044301 (2003).
\bibitem{denisov2003}
S. Denisov,  J. Klafter and  M. Urbakh,  Phys. Rev. Lett. \textbf{91},  194301 (2003).
\bibitem{cipriani2005}
P. Cipriani,  S. Denisov and  A. Politi,  Phys. Rev. Lett. \textbf{94},  244301 (2005).
\bibitem{libw2005}
B. Li,  J. Wang, L. Wang and  G. Zhang,  Chaos \textbf{15}, 015121 (2005).
\bibitem{zhao2006}
H. Zhao, Phys. Rev. Lett. \textbf{96}, 140602 (2006).
\bibitem{delfini2007}
L. Delfini, S. Denisov, S. Lepri,  R. Livi,  P.K. Mohanty and A. Politi,  Eur. Phys. J. Special Topics
 \textbf{146}, 21 (2007).
\bibitem{einstein1905}
A. Einstein,  Annalen der Physik \textbf{322},  549 (1905).
\bibitem{mai2007}
T. Mai,  A. Dhar and  O. Narayan,  Phys. Rev. Lett. \textbf{98},  184301 (2007).
\bibitem{delfini199401comm}
L. Delfini,  S. Lepri,  R. Livi and  A. Politi,  Phys. Rev. Lett. \textbf{100},  199401 (2008).
\bibitem{dhar2008Re}
A. Dhar and  O. Narayan,  Phys. Rev. Lett. \textbf{100},
  199402 (2008).
\bibitem{zavt1993}
G.S. Zavt,  M. Wagner and  A. L{\"u}tze,  Phys. Rev. E \textbf{47},  4108 (1993).
\bibitem{arevalo2003}
E. Ar{\'e}valo,  F.G. Mertens,  Y. Gaididei and  A.R. Bishop,  Phys. Rev. E \textbf{67},  016610 (2003).
\bibitem{kopidakis2008}
G. Kopidakis,  S. Komineas,  S. Flach and  S. Aubry,  Phys. Rev. Lett. \textbf{100},  084103 (2008).
\bibitem{dhar2008}
A. Dhar and  J.L. Lebowitz,  Phys. Rev. Lett. \textbf{100},  134301 (2008).
\bibitem{marconi2008}
U.M.B. Marconi,  A. Puglisi,  L. Rondoni and  A. Vulpiani,  Phys. Rep. \textbf{461},  111 (2008).
\bibitem{ponno2006}
A. Ponno,  J. Ruggiero,  E. Drigo and  J.DeLuca,  Phys. Rev. E
\textbf{73},  056609 (2006).
\bibitem{zhao2005}
H. Zhao,  Z. Wen,  Y. Zhang and  D. Zheng,  Phys. Rev. Lett. \textbf{94},  025507 (2005).
\end{thebibliography}
\end{document}